\documentclass[aps,prl,twocolumn,superscriptaddress,groupedaddress]{revtex4}  
\usepackage{epsfig}
\usepackage{graphicx}  
\usepackage{dcolumn}   
\usepackage{amssymb}
\usepackage{physics}
\usepackage{natbib}
\usepackage{color}
\usepackage{lineno}
\usepackage{float}

\begin{document}

\title{
Phase-Slip Statistics of a Single Isolated Flux-Biased Superconducting Ring
}

\author{I.~Petkovic }
\affiliation{Department of Physics, Yale University, 217 Prospect Street, New Haven, Connecticut 06520, USA}
\author{A.~Lollo}
\affiliation{Department of Physics, Yale University, 217 Prospect Street, New Haven, Connecticut 06520, USA}
\author{J.~G.~E.~Harris}
\affiliation{Department of Physics, Yale University, 217 Prospect Street, New Haven, Connecticut 06520, USA}
\affiliation{Department of Applied Physics, Yale University, 15 Prospect Street, New Haven, Connecticut 06520, USA}

\begin{abstract}
We describe measurements of the thermally activated transitions between 
fluxoid states of a single isolated superconducting ring. We compare
these measurements with theoretical predictions in which all of the
relevant parameters are determined via independent characterization of
the same ring. This no-free-parameters comparison shows qualitative
agreement over a wide range of temperatures. We discuss possible origins
for the remaining discrepancies between the data and theory, in
particular the choice of model for the superconducting order parameter's
damping.
\end{abstract}

\maketitle

A current-carrying state of a one-dimensional (1D) superconductor is described within Ginzburg-Landau (GL) theory as metastable, i.e., occupying a local minimum of the free-energy landscape \cite{lan,mhh}. In such a state, the position-dependent phase is characterized by a winding number \cite{little1,little2}.
A phase slip is a change of the winding number and thus a passage from one metastable minimum to another, across a barrier, that is overcome either by thermal activation or by quantum tunneling \cite{lan,mhh}.
The free energy of metastable states and the barrier heights were calculated in the pioneering work of Langer and Ambegaokar \cite{lan}, and McCumber and Halperin \cite{mhh}. McCumber and Halperin calculated the escape rate for thermal activation across the barrier $\Gamma=\omega_a \exp(-\delta F/k_B T)$, with $\omega_a$ the attempt frequency, $\delta F$ the free-energy barrier, $T$ the temperature, and $k_B$  the Boltzmann constant \cite{mhh}. This entire result is typically referred to as the Langer-Ambegaokar-McCumber-Halperin (LAMH) theory. However, in order to distinguish between the stationary properties (free energy of extremal states) and nonstationary properties (escape rates), we will refer to the former as 1D GL and the latter as LAMH. 

Better quantitative understanding of thermal phase slips is a prerequisite for the study of incoherent \cite{giordano_1,giordano_2,giordano_3} and coherent \cite{mooij_nazarov} quantum phase slips, the latter of which could be used as a basis for quantum computation \cite{mooij_harmans}. It could also further elucidate studies of superconducting fluctuations above the critical temperature \cite{koshnick,schwiete_oreg,vonoppen_riedel}, superconducting-insulating transitions \cite{buchler,hekking_glazman}, quasiparticle dynamics \cite{rob_qp,michel_qp,kamenev_qp}, topological order in hybrid structures containing superconducting elements \cite{roman,majorana_vonoppen}, and superconducting qubit decoherence \cite{michel_rob_outlook}.

Experimental studies of phase slips have focused primarily on either current-biased superconducting wires or flux-biased isolated rings. 
In wires, the phase-slip rate gradually increases as the bias current is increased up to the critical current, its only instability point. A phase slip can occur at any value of the bias with some probability, and each such event produces a tiny voltage pulse \cite{lan,mhh}. In practice, an individual phase slip is very difficult to detect. Early experiments were indirect, measuring the cumulative effect of voltage pulses as an effective resistance \cite{newbower,giordano_1,lau_qps,bezryadin_qps,altomare,zgirski,bollinger,rogachev_1,rogachev_2}. In some experiments the heating due to a single phase slip could drive the whole wire to a normal state; this effect was used to isolate individual phase slips and measure their statistics \cite{bezryadin_sahu,li,aref}. However, this presented a limitation on the applicable temperature range, and the study is complicated by the effects of heating.

In the case of a flux-biased ring, the winding number denotes the number of fluxoid quanta in the ring \cite{byers_yang,little1,little_parks2}. For each winding number, a phase slip always occurs in the close vicinity of the flux value at which the metastable state becomes unstable as persistent current reaches the critical current $I_{\rm c}$. The phase slip is accompanied by a jump in the persistent current of the order of $I_{\rm c}$, which is easily detectable with the appropriate contactless measurement technique.
The ring geometry is thus amenable to isolating individual phase-slip events and was used to study their statistics \cite{belkin, ps_rings_1,ps_rings_2,zhang_price,budakian}. In the ring geometry the whole system does not transition into the normal state during a phase slip; therefore, the phase slip can be measured in the full temperature range, and the effects of heating become negligible. Since the rings are not connected to an external circuit, one is in principle able to access their intrinsic phase dynamics. This isolation also simplifies the boundary conditions for the theoretical analysis of the problem.  

The usual procedure in both configurations was to fit measured escape rates with LAMH theory and check whether the inferred system parameters (such as coherence length or penetration depth) were plausible for the system in question, usually without the possibility of independent verification. Since the escape rate is exponential with respect to the free energy, a small variation in the system parameters  (well within the plausibility range) leads to a large change in the calculated escape rates. Therefore, independent access to system parameters is necessary in order to study the escape rate in finer quantitative detail. 

In order to quantitatively study the escape rate, it would be advantageous to characterize the system parameters via independent measurements. In previous work \cite{deterministic} we characterized an ensemble of $\sim 10^2$ flux-biased isolated rings by measuring their persistent current as function of flux $I(\phi)$ at various temperatures and fitting the result with 1D GL theory. This enabled us to extract the system parameters (as fit parameters): ring radius $R$, superconducting coherence length $\xi$, penetration depth $\lambda$, and ring width $w$. We found that phase slips occur close to critical flux values $\phi_{\rm c}$, at which the barrier between metastable states $\delta F$ is predicted to vanish by 1D GL. However, it was not possible to measure thermally activated switching in these samples owing to the inhomogeneity of the ensemble. Here we measure a single ring and study the statistics of its thermally driven escape from metastable states.

The persistent current $I$ of an isolated flux-biased superconducting ring is detected via its magnetic moment $\mu=I R^2 \pi$ using cantilever torque magnetometry \cite{will_thesis}. The system, shown in Fig. 1 in the Supplemental Material (SM) \cite{si}, consists of an aluminum ring atop a silicon microcantilever placed in an external perpendicular magnetic field $B$. At rest, $\mu$ is collinear with the applied field, and there is no torque. However, as the cantilever oscillates, $\mu$ exerts a torque on the cantilever, thus changing its resonant frequency by an amount $df = \kappa I(B) \cdot B$, where $\kappa$ is a constant that contains the spring constant $k$ of the cantilever. The detailed sample fabrication and measurement principle are described elsewhere \cite{will_thesis}.

\begin{figure}[h!]
\centerline{
\hbox{
\epsfig{figure=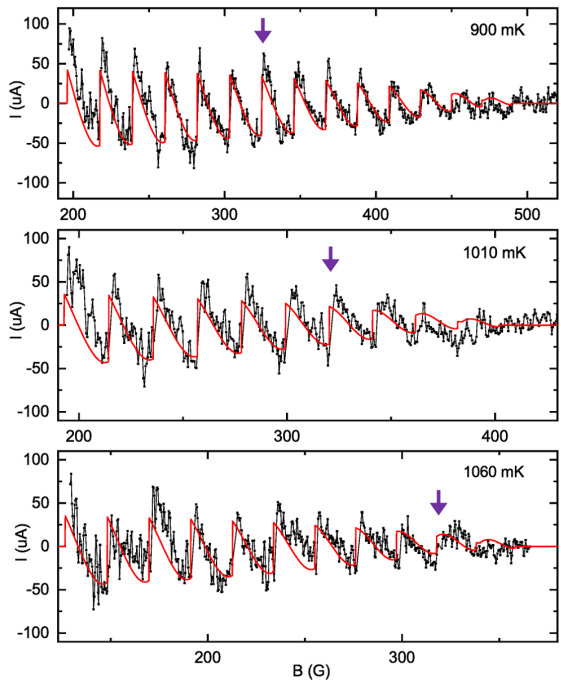,width=75mm}}}
\caption{$I(B)$ measured for increasing $B$ (black) and fit (red) for a ring with $R=546.8$ nm and $T=900, 1010$, and $1060$ mK top to bottom. The purple arrow denotes the phase slip between winding numbers 12 and 13, on which the thermal activation is subsequently measured. 
}
\label{f1}
\end{figure}

We start by characterizing the single ring in the stationary states. Figure \ref{f1} shows measured $I(B)$  for a single ring of $R=546.8$ nm at various temperatures. Data is shown as black curves, and the 1D GL fit is given as red curves. The fit follows \cite{zhang_price} and includes the effect of the field penetration into the superconductor (also referred to as the finite-width effect), which causes the persistent current to decay with increasing field. 
For materials with short mean free path, the validity of GL is extended from the immediate vicinity of $T_{\rm c}$ to lower temperature, and in previous work we found that the validity range for the aluminum rings used is $T>750$ mK (see \cite{deterministic} and discussion therein) or above $T_{\rm c}/2$, with $T_{\rm c}=1.3$ K. 
The fitting procedure is detailed in the SM \cite{si}.

We find as fitting parameters the zero temperature coherence length $\xi_0 \sim 210$ nm, zero temperature penetration depth $\lambda_0 \sim 110$ nm, and the ring width $w=64$ nm. These values are consistent with those obtained on an ensemble of nominally identical rings \cite{deterministic}.


\begin{figure}[h!]
\centerline{\hbox{
\epsfig{figure=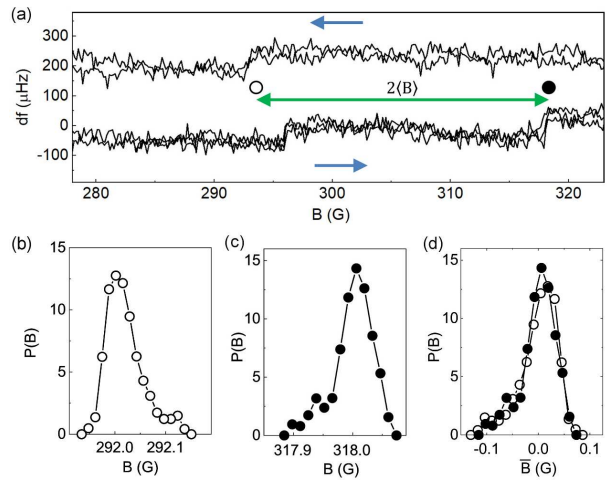,width=80mm}}}
\caption{(a) Measurement of $df(B)$ over the two phase slips that were repeatedly swept in order to collect the histograms, with blue arrows as sweep directions. The sweep-down traces are shifted vertically for clarity. The location of the up/down jump is denoted with a full/empty black dot. The distance between the up and down jump  is denoted as 2$\langle B \rangle$ (green arrow).  The absence of a jump around 310 G for sweep down is due to the size of the histogram loop.  (b),(c) Histograms for sweep down/up (empty/full dots, respectively). (d) The sweep-down histogram is flipped around the mean and superimposed on the sweep-up histogram. Both histograms are shown around zero mean. These measurements were performed at 700 mK. 
}
\label{f2}
\end{figure}

\begin{figure*}
\centerline{
\hbox{
\epsfig{figure=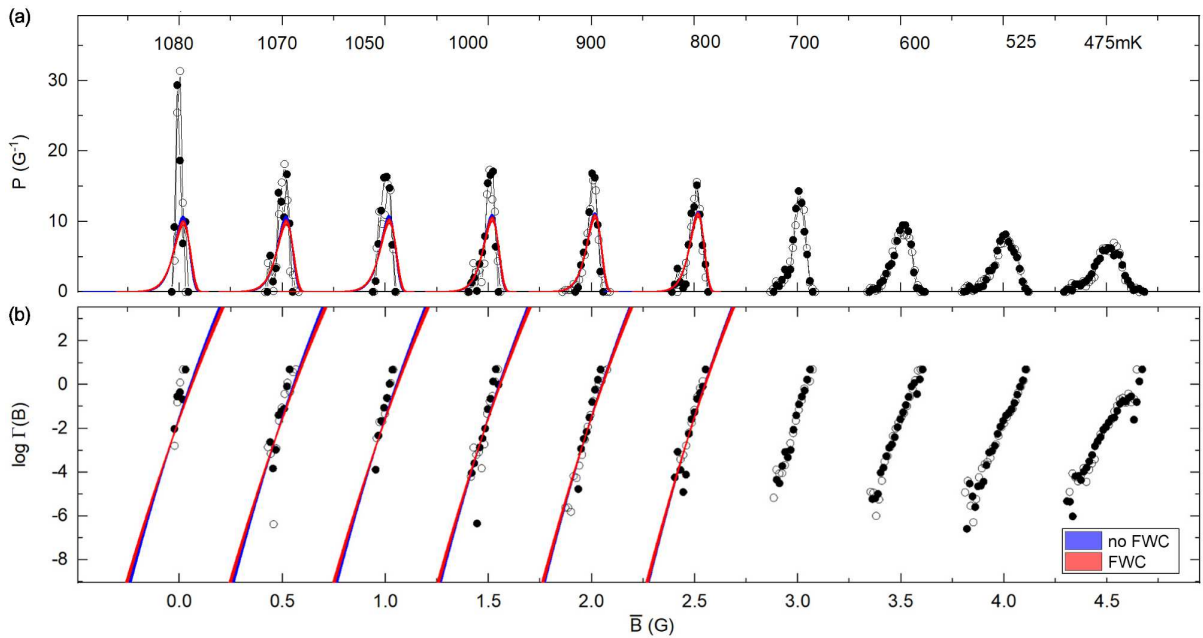,width=160mm}}}
\caption{(a) Histograms as a function of temperature. Circles are data, full symbols denote sweep up and empty sweep down. Temperature is  marked above each histogram. Histograms are shifted horizontally for clarity. (b) Escape rates obtained from the histograms in (a). Both  panels: colored regions are the LAMH prediction for histograms/escape rate respectively in the two regimes, no finite-width correction (FWC) (blue) and FWC (red). }
\label{f3}
\end{figure*}

After characterizing the ring over the full range of $B$ (as shown in Fig. \ref{f1} of the main text and Fig. 3 of the SM \cite{si}), we now focus on detailed measurements of the narrow range of $B$ corresponding to the phase-slip transitions between winding numbers 12 and 13.
As shown in Fig. \ref{f2}(a), a small loop in $B$ was repeatedly swept, and values at which the jumps occur  (denoted by full/empty circle for jump up/down respectively) were recorded. The ramp rate was 13 mG/s, and each histogram contains several hundred events. The data was binned and shown as histograms [Figs. \ref{f2}(b)-(d)] where
sweep-up histograms are shown as full circles and sweep-down histograms are flipped and shown as empty circles.   This particular phase slip disappears at $T \approx 1.1$ K.  

\begin{figure}
\centerline{
\hbox{
\epsfig{figure=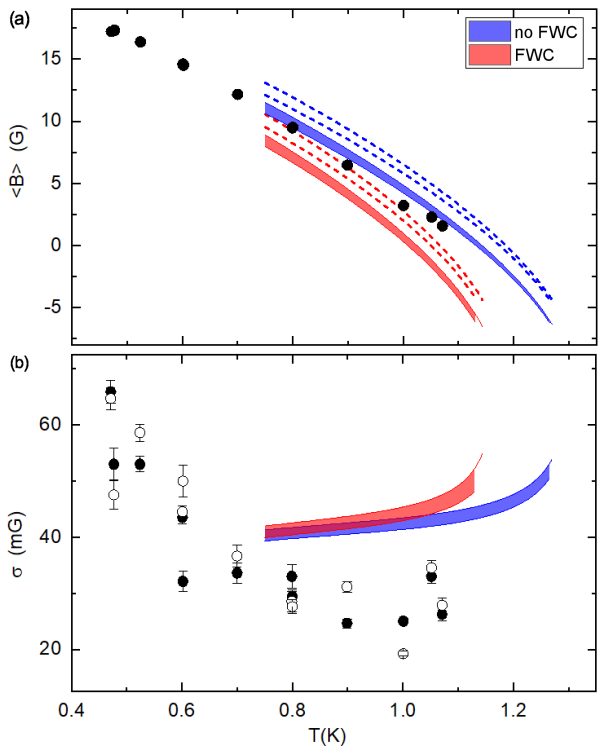,width=80mm}}}
\caption{(a)  Mean switching value $\langle B(T) \rangle$, data are shown as circles. (b) Histogram width $\sigma(T)$, where full/empty dots are data for sweep-up/down histograms. Dotted lines in (b) are a prediction for $B_{\rm c}$. Both panels: colored regions are the LAMH prediction for mean/width respectively in the two regimes, no FWC (blue) and FWC (red), corresponding to cases with or without the finite-width correction, respectively.  }
\label{f4}
\end{figure}

In Fig. \ref{f3}(a) we show the resulting histograms. 
The same data is converted to escape rates following Eq. (6) in \cite{fulton_dunkelberger} and shown in Fig. \ref{f3}(b).
From the histograms we extract the mean switching value $\langle B \rangle = (\langle B \rangle_{\rm up} - \langle B \rangle_{\rm dw})/2$, shown as function of temperature in Fig. \ref{f4}(a). Here $\langle B \rangle_{\rm up,dw}$ are mean values of the phase-slip field in the up/down sweep direction respectively. We also extract the histogram standard deviation $\sigma$ shown in Fig. \ref{f4}(b) as full/empty dots for the up/down sweep direction. The error bars for $\sigma$ stem from uncertainty introduced by the background removal \cite{si}. 

Next we compare the measured histograms to the LAMH theory. We emphasize that at this point there are no more fitting parameters, since we use the values for $\xi$, $\lambda$, and $w$ obtained from the 1D GL fit and simply plug them into the escape rate expressions in \cite{mhh}, with the prefactor given by Eq. (4.35a) in \cite{mhh}. 
We include the finite-width correction 
by noticing that the free-energy barrier $\delta F(B)$ can be expressed in terms of the critical current and persistent current \{Eq. (3.23) in \cite{lan}\}, and we scale these quantities using the 1D GL values calculated previously. In Fig. \ref{f3} we show both cases, without (blue) and with the finite-width correction (red). A region of one color covers the values of $\xi\pm\delta \xi$ and $w \pm \delta w$ where $\delta \xi$ and $\delta w$ are error bars to the fit parameters. The difference between calculated histograms/escape rates for the two cases is very small.  
Data is shown in the full accessible temperature range and theory in the GL applicable range.

Qualitatively, we expect $\langle B \rangle \sim B_{\rm c}^n$, where $B_{\rm c}^n = \phi_{\rm c}^n \Phi_0/R^2 \pi$ is the critical value of field such that $\delta F_n(B_{\rm c}^n)=0$ for the winding number $n$. (Here $\phi$ is reduced flux, with $\Phi_0$ the superconducting flux quantum.) In Fig. \ref{f4}(a) we compare the measured $\langle B \rangle$ (black dots) to the LAMH prediction, shown as a blue/red region. The dashed lines show the theoretical prediction for $B_{\rm c}$, again for the range $\xi \pm \delta \xi$ and $w \pm \delta w$.  We see that the theory roughly reproduces the measurement, either with or without the finite-width correction.

In Fig. \ref{f4}(b) we compare the measured histogram widths $\sigma$ to the LAMH theory. We see that the measured values are systematically smaller than the prediction. However, the difference between theory and prediction is comparable to the scatter in the data. The more surprising aspect of the data is that the histograms get broader as the temperature is lowered. This is the reverse from what is expected for purely thermal fluctuations with constant damping \cite{fulton_dunkelberger,garg}. 
We note that low temperature broadening of histograms has been observed on the phase-slip statistics of current-biased wires using the previously mentioned effect of heating that drives the wire into the normal state \cite{bezryadin_sahu,li}. In \cite{bezryadin_sahu} the broadening was attributed to a combination of three factors: multiple phase slips at higher bias, heating, and quantum phase slip contribution at lower temperatures. 
Here, we can exclude all three contributions. Multiple phase slips are excluded because the  directly measured height of the $I(B)$ jump and the spacing between jumps correspond to a single flux quantum in all of the measurements presented here. Quantum phase slips are excluded because the ring's resistance $\sim 10$  $\Omega$ (based on a measurement of the mean free path of an aluminum wire on the same chip), well below the quantum of resistance for electron pairs \cite{buchler}. Heating is expected to play a much smaller role in the present experiment than in \cite{bezryadin_sahu} due to the size of the rings, their coupling to the substrate, and the fact that the system never transitions to a resistive state.

Now we aim to analyze the difference between experimental escape rates and those predicted by theory [Fig. \ref{f3}(b)]. We assume that the experimental escape rate has an additional factor $\eta$ in the exponent and can be written in the form

\vspace{-3mm}

\begin{equation}
\Gamma =\omega_a  \exp{- \frac{\eta \, \delta F(B)}{k_B T}}.
\end{equation}

\vspace{1mm}

\begin{figure}[h!]
\centerline{
\hbox{
\epsfig{figure=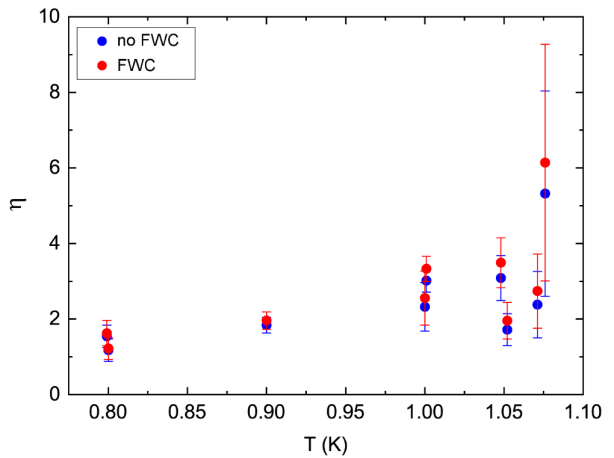,width=80mm}}}
\caption{The factor $\eta$ in the escape rate exponent [see Eq. (1)] extracted as detailed in the text. } 
\label{f5}
\end{figure}

\noindent In LAMH we have $\eta=1$, and therefore the deviation of the experimental values from theory would result in $\eta \neq 1$. Since $\log \Gamma \sim \eta \, \delta(B)/k_B T$ and, in the small range of $B$ covered by the measurement, $\log \Gamma$ is a  nearly linear function of bias, we can calculate $\eta$ as the ratio of the slopes of the linear fits to the measured $\log \Gamma$ and the prediction of Eq. (1) (for $\eta=1)$, i.e., between the dots and the line shown in Fig. \ref{f3}(b) for each temperature. Since we have two cases for the predicted $\Gamma$ [with (red) or without the finite-width correction (blue)], the $\eta$ values are also calculated for those two cases. The obtained result, $\eta(T)$, is shown in Fig. \ref{f5}. We see that in the full range $\eta \gtrsim 1$ and that $\eta$ increases as the temperature goes up. 

One possible explanation for $\eta \gtrsim 1 $ is that the free-energy barrier is somewhat higher than predicted by LAMH. In this work we use the LAMH theory valid for an infinitely long wire/ring, while we have shown in previous work that for a finite-length ring the phase slip actually occurs beyond $\phi_{\rm c}$, at a higher value $\phi^*=\phi_{\rm c}+O(R/\xi)$ \cite{deterministic}. This stems from the Eckhaus nonlinearity of GL equations  \cite{eckhaus,khlebnikov}. The calculation for the free-energy barrier in this case is not available (within GL).

It is also known that a temperature-dependent $\eta$ can result if the thermal activation takes place in the presence of damping that is not simply a constant (e.g., if the damping is frequency dependent or takes on different values at the free-energy minima and maxima) \cite{ambegaokar_halperin,martinis_kautz,michel1,michel2}. In the present experiments, we believe the ring's dynamics are overdamped \cite{deterministic} (consistent with the assumptions of LAMH), since we only observe the flux to change by a single quantum, but we do not have a more detailed model of the damping from which to estimate $\eta(T)$. We note that such a model could be developed by extending the results of Ref. \cite{boogard}.

In conclusion, we have measured the escape rates of phase slips in a single, isolated, homogeneous, one-dimensional superconducting ring. We have characterized the system by fitting the $I(B)$ data over the full superconducting range with 1D GL and then used the obtained system parameters to calculate the LAMH escape rates.
We compare the resulting prediction to the measured escape rates without any free parameters and find rough agreement with the LAMH prediction. The remaining discrepancy is expressed via a factor $\eta$ in the escape rate exponent, where we find $\eta \gtrsim 1$. We discuss the possible provenance of this factor from two sources. One is the refinement of the calculation of the free-energy barrier to include the stabilization beyond the critical point, characteristic of the Eckhaus instability of 1D GL. The second is that $\eta$ may rise from damping that varies with frequency or within the free-energy landscape, as has been observed for Josephson junctions in the phase diffusion regime.  

We would like to thank Michel Devoret, Leonid Glazman, Bert Halperin, and Teun Klapwijk for useful discussions. We also thank Ania Bleszynski-Jayich and Will Shanks for sample fabrication. We acknowledge support from the National Science Foundation, Grant No. 1106110.

\clearpage

\onecolumngrid
\begin{center}
\vspace{4mm}
\large{\textbf{ Supplemental Material for "Phase-Slip Statistics of a Single Isolated Flux-Biased Superconducting Ring" }}
\end{center}

\section{Sample}

\setcounter{figure}{0}

Figure 1 shows the SEM photo of the sample. The silicon microcantilever with a single Al ring on top is shown in the left panel, and the zoom in on the ring is shown in the right panel. The fabrication is described in \cite{will_thesiss}.

\begin{figure}[h!]
\centerline{
\hbox{
\epsfig{figure=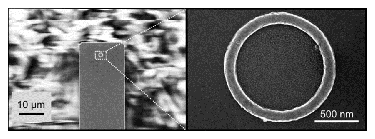,width=150mm}}}
\caption{SEM photo of the sample. Left: silicon microcantilever with one single aluminum ring such as the one used in the experiment. Right: zoom in on the ring. From \cite{will_thesiss}.
}
\label{f1sm}
\end{figure}

\section{ $I(B)$ fit }

The conversion between the measured frequency shift $df$ and the persistent current $I$ depends on the relative angle between the sample's magnetic moment $\mu$ and the applied field $B$. Here, they are parallel when the cantilever is at rest, creating a torque as it oscillates, and $df=\kappa I(B) \cdot B$, where $\kappa$ is a constant which contains the spring constant $k$ of the cantilever.  We see that signal is low around zero field, as the torque on the cantilever is very small, despite the fact that $I(B)$ itself is the highest in that range. This measurement configuration is not appropriate for the measurement of persistent current when $B \rightarrow 0$. That is the reason that we could not study the $n=0$ phase slip. 

\subsection{Fitting procedure}

The fitting procedure is similar to that of our previous work \cite{deterministicc}. Here we repeat the main parts for completeness and focus on the differences. 

The global fit of $I(B)$ is done for fitting parameters $\xi(T)$, penetration depth $\lambda(T)$, ring radius $R$, ring width $w$ and cantilever spring constant $k$, using Eq. (7) from \cite{zhang_pricee}. Some of these parameters are expected to be temperature-dependent and some are not. Therefore the fit is conducted in two rounds. In the first round all parameters are free, and the fitting is performed for each temperature. Then we fix $k$ and $w$ at the means obtained by these fits, and carry out the second round where the only free parameters are $\xi$ and $\lambda$ at each value of $T$. $R$ can be obtained from the fit, but it can also be obtained to high accuracy by fitting only jump locations, separately before starting the fit, and this is done as it is more practical. 

The main difference for a single-ring measurement, as compared to the ensemble in \cite{deterministicc}, is that the signal is small, which has several consequences. The background is now comparable to the signal and the true background is not a priori known. We therefore remove the background so as to make each phase slip jump symmetric around zero. We remove the background from the fit function as well - this is done by calculating the background for initial values of the fit parameters, removing it, and then fitting, and in the end checking that the obtained values are not too far from the initial values. The critical field $B_{\rm c3}$ cannot be read out directly from the data (unlike for the ensemble, \cite{deterministicc}). We therefore include it as a fitting parameter. Note that $\xi$, $w$ and $B_{\rm c3}$ are not independent, since $B_{\rm c3} \sim 1/(\xi w)$. This limits the range of initial values of $B_{\rm c3}$ to the plausible values of $w$, the range of which is read out from the SEM image of the sample. We use the value of $T_{\rm c}$ obtained for the ensemble \cite{deterministicc}, since both samples are on the same chip and were fabricated together. We use the finding of the previous paper that the phase slip occurs at $B_{\rm c} \sim \phi_{\rm c}$ and we fit the data to the full curve, with included jumps.

\subsection{Ring $R=546.8$ nm}

In Fig. \ref{f2} we show the measured $df(B)$ in the full temperature range for two sweep directions (denoted by black arrows). This measurement is conducted on a single ring of radius $R=546.8$ nm.  The slow random background is removed such that jumps in $df$ are symmetric around the line $df=0$. Each jump is a phase slip, i.e., a change in the winding number of the phase of the superconducting order parameter. The periodicity of the sawtooth pattern is due to the changing number of flux quanta that pierce the ring. Each jump is a change by a single flux quantum, or a change of the winding number by one. Even at the lowest accessible temperature we do not observe jumps by multiple flux quanta. We denote the applied field at which persistent current goes permanently to zero $B_{\rm c3}$ and we see that $B_{\rm c3}$ is diminished as temperature is increased. 

\vspace{3mm}

\begin{figure}[h!]
\centerline{
\hbox{
\epsfig{figure=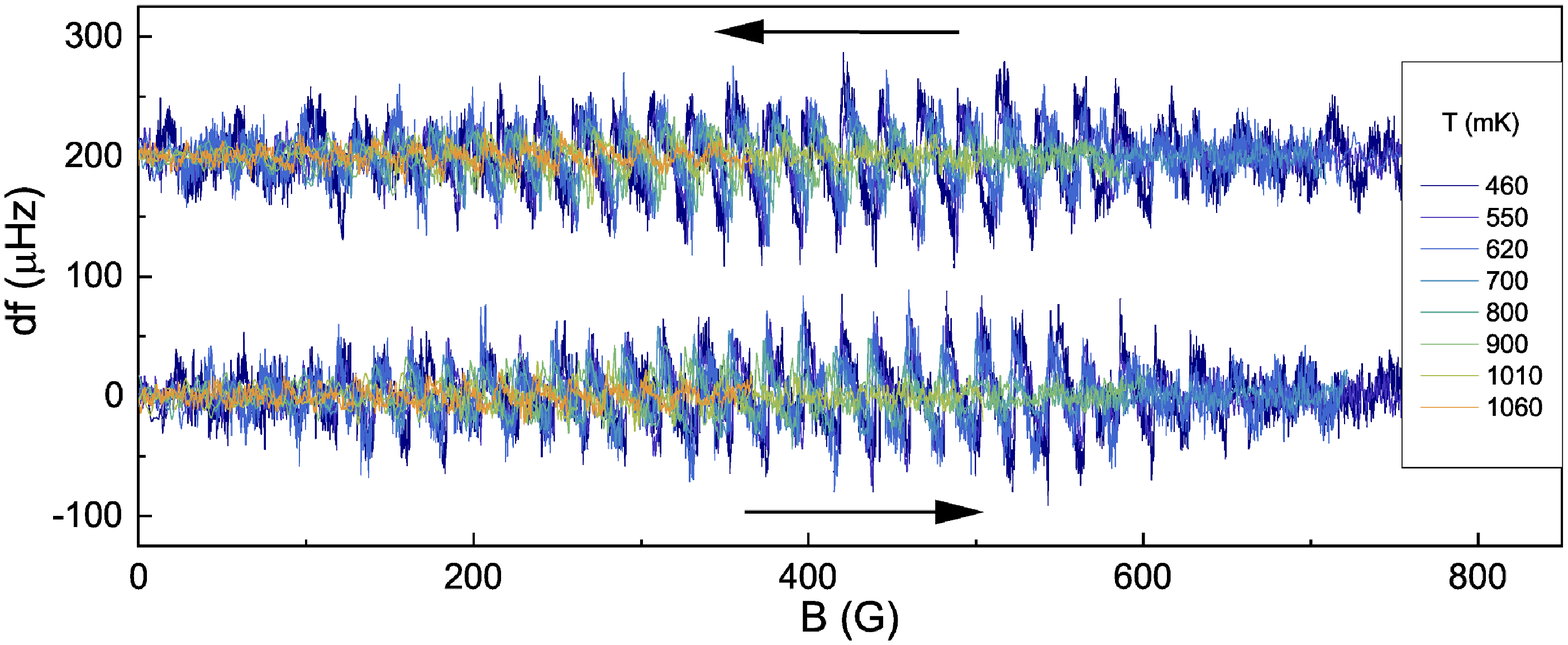,width=150mm}}}
\caption{$df(B)$ measured for the ring with $R=546.8$ nm in the full range of temperature denoted in the legend. Sweep direction is denoted by black arrows. In this measurement configuration $df(B) \sim I(B)\cdot B$. 
}
\label{f2sm}
\end{figure}

In Fig. \ref{f3} we show this same dataset in separate panels as black dots, but now converted from $df(B)$ to $I(B)$ and focusing on the temperature range $T>750$ mK, where we fit the data with the 1D GL (or the stationary part of LAMH) theory. The version of the theory used here takes into account the finite ring width, and is detailed in \cite{zhang_pricee}. The fit is shown as red curves. Each trace between $-B_{\rm c3}$ and $B_{\rm c3}$ is split into two traces, one from $0$ to $B_{\rm c3}$, and another from $-B_{\rm c3}$ to $0$. The latter is transformed along $B \rightarrow -B$ and $I \rightarrow -I$, which we refer to as ,,flipped". Because of symmetry, ,,up" and ,,down flipped" should be identical, as should ,,down" and ,,up flipped", since $I(-B)=-I(B)$. We use this redundancy to fit these traces separately and obtaining very similar result is a common sense check on our method.

\begin{figure}[h!]
\centerline{
\hbox{
\epsfig{figure=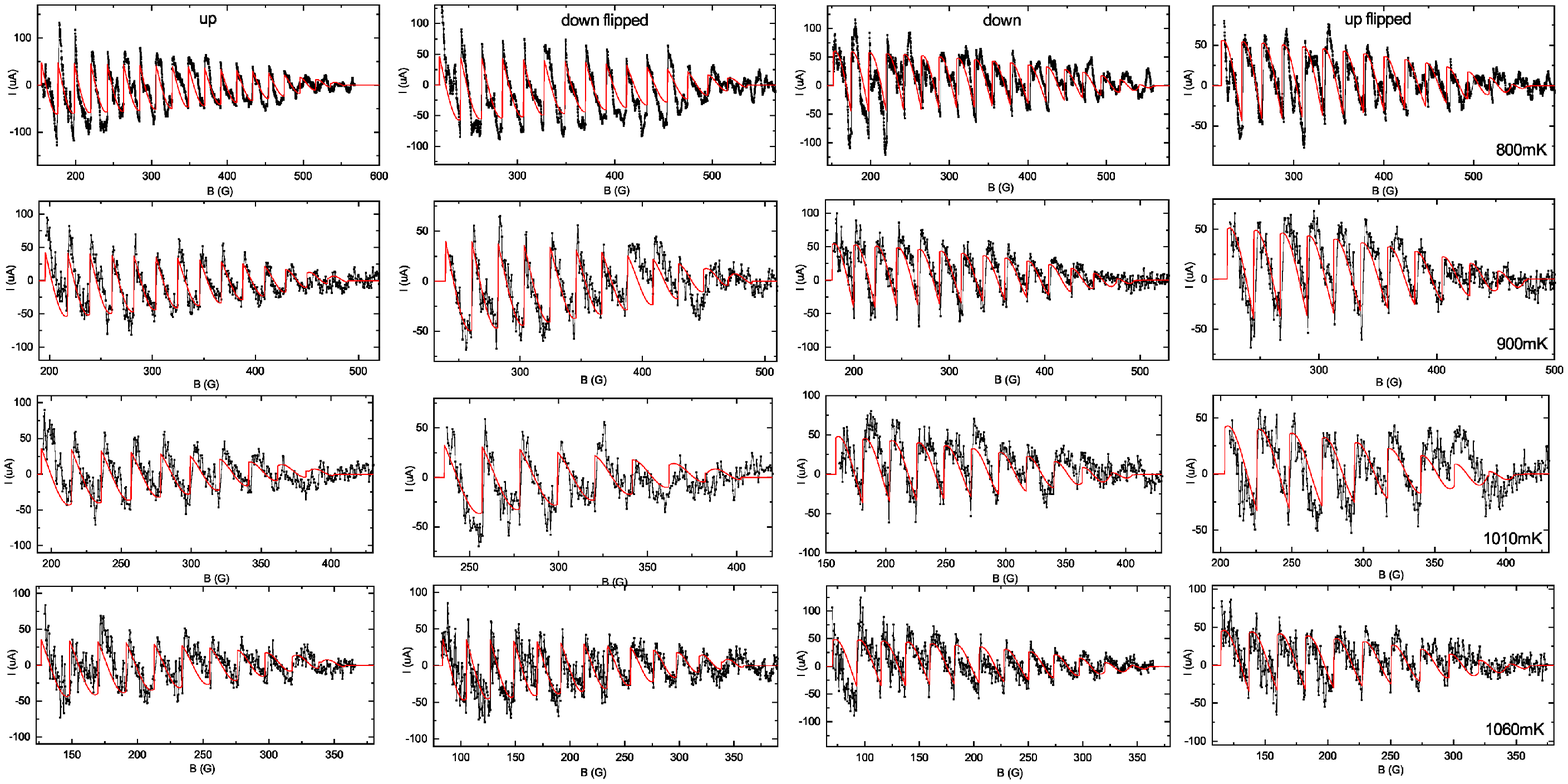,width=180mm}}}
\caption{$I(B)$ measured (black) and fit (red) for the ring with $R=546.8$ nm and $T=800,900,1010$, and $1060$ mK top to bottom in rows. Columns are sweep directions: up, down flipped, down and up flipped. Down flipped is obtained by taking the portion of the down sweep $I(B)$ with $B<0$ and flipping $B \rightarrow -B$ and $I \rightarrow -I$. Equally up flipped is obtained by taking the portion of the up sweep $I(B)$ with $B<0$ and flipping $B \rightarrow -B$ and $I \rightarrow -I$. Due to symmetry up and down flipped should coincide, as should down and up flipped, since $I(-B)=-I(B)$. They are given separately for clarity.  
}
\label{f3sm}
\end{figure}

Each trace gives as fitting parameters $\xi$ and $B_{\rm c3}$ at that temperature, and in Fig. \ref{f4} we show all obtained $\xi$ and $B_{\rm c3}$ values and check that they follow the temperature dependence as expected. This is another check on the overall method. We fit $\xi$ and $B_{\rm c3}$ (fit given in blue) with 

$$
\xi(T)=\xi_0 \sqrt{\frac{1+\left( \frac{T}{T_{\rm c}}\right) ^2}{1-\left( \frac{T}{T_{\rm c}}\right) ^2}} \hspace{16mm} B_{\rm c3} = B_{\rm c3,0} \sqrt{\frac{1-\left( \frac{T}{T_{\rm c}}\right) ^2}{1+\left( \frac{T}{T_{\rm c}}\right) ^2}}
$$ 

\vspace{3mm}

\begin{figure}[h!]
\centerline{
\hbox{
\epsfig{figure=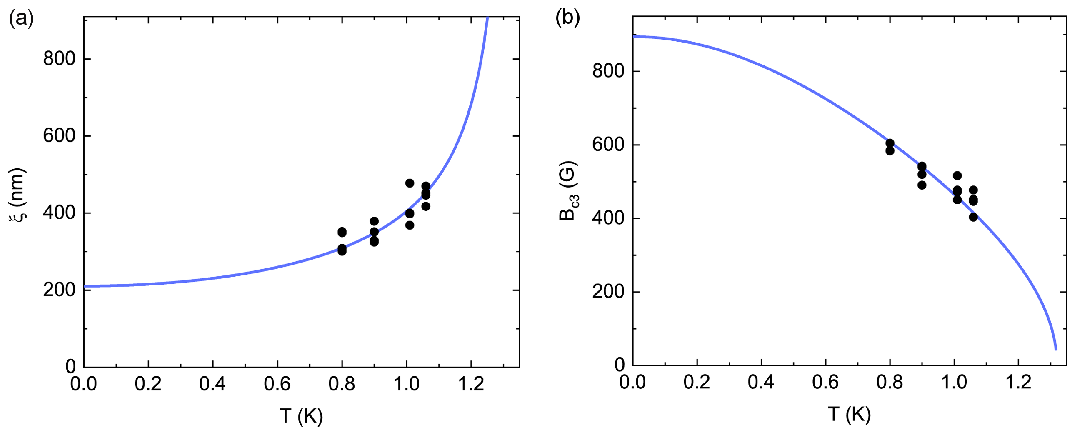,width=150mm}}}
\caption{(a) $\xi(T)$: black dots are values obtained as fit parameters in $I(B)$ and blue curve is the theory curve with $\xi_0=210.6$ nm and $T_{\rm c}=1.32$ K. (b) $B_{\rm c3}$: black dots are values obtained as fit parameters in $I(B)$ and blue curve is the theory curve with $B_{\rm c30}=894.61$ G and $T_{\rm c}=1.32$ K.
}
\label{f4sm}
\end{figure}

\noindent We find $\xi_0=210.6$ nm and $B_{\rm c3,0}=894.6$ G.

The parameter values obtained for a single ring are consistent with those obtained on an ensemble of nominally identical rings. 
Both the ensemble  and the single ring studied here were fabricated simultaneously on the same chip.
Since both $w$ and the ring thickness ($d=90$ nm) are smaller than $\xi_0$ and $\lambda_0$, it is confirmed that the 1D GL theory is applicable.

\section{Histograms }

Sweeps to collect the histograms were carried out with the main solenoid in the persistent mode at a fixed field value. 
The magnetic field was swept by varying the current in a smaller (2.5 cm diameter) solenoid which was placed around the sample holder, and above the piezo element that actuated the cantilever. Current lines to the small coil were filtered. 
The data set showed a close-to-linear slow background, which was subtracted from the data. This background likely originated from the slow decay of current from the main solenoid.

\section{Asymptotic behavior of $\delta F$ near $B_{\rm c}$}

If we use the notation from the Langer-Ambegaokar paper \cite{la}, we can write 

$$
\delta F = \frac{\hbar}{e}\sqrt{\frac{3}{2}} I_{\rm c} \left( \sqrt{\Delta} - \sqrt{\frac{2}{3}}\frac{I}{I_{\rm c}} \arctan \sqrt{\frac{\Delta}{2 (1-\Delta)}}        \right)
$$

\noindent where

$$
(2+\Delta)^2 (1-\Delta)= 4 \left(  \frac{I}{I_{\rm c}} \right)^2.
$$

In the limit $I \rightarrow I_{\rm c}$ we have $\Delta \rightarrow 0$ and 

$$
\delta F \sim \frac{\hbar}{e}\sqrt{\frac{3}{2}} I_{\rm c} \Delta^{\frac{5}{2}} \sim (B_{\rm c} - B)^{\frac{5}{2}}. 
$$

\noindent We give this exponent of $5/2$ explicitly since it is characteristic of phase slips in one dimension. If we had a case of phase slip not in a uniform ring, but with a weak link somewhere along the ring, the exponent would be different.


\begin{thebibliography}{99}

\bibitem{lan} J.S. Langer and V. Ambegaokar,  Phys. Rev. \textbf{164}, 498 (1967).

\bibitem{mhh} D.E. McCumber and B.I. Halperin, Phys. Rev. B \textbf{1}, 1054 (1970).

\bibitem{little1} W.A. Little, Phys. Rev. \textbf{134}, A1416 (1964).


\bibitem{little2} W.A. Little,  Phys. Rev. \textbf{156}, 396 (1967).





\bibitem{giordano_1} N. Giordano, Phys. Rev. Lett. \textbf{61}, 2137 (1988).

\bibitem{giordano_2} N. Giordano and E.R. Schuler, Phys. Rev. Lett. \textbf{63}, 2417 (1989)

\bibitem{giordano_3} N. Giordano, Phys. Rev. B \textbf{41}, 6350 (1990).

\bibitem{mooij_nazarov} J.E. Mooij and Yu.V. Nazarov, Nat. Phys. \textbf{2}, 169 (2006).

\bibitem{mooij_harmans}  J.E. Mooij  and C.J.P.M. Harmans, New J. Phys. \textbf{7}, 219 (2005).

\bibitem{koshnick} N.C. Koshnick, H. Bluhm, M.E. Huber, and K.A. Moler, Science \textbf{318}, 1440 (2007).

\bibitem{schwiete_oreg} G. Schwiete and Yu. Oreg, Phys. Rev. B \textbf{82}, 214514 (2010).

\bibitem{vonoppen_riedel} F. von Oppen and E.K. Riedel, Phys. Rev. B \textbf{46}, 3203(R) (1992).

\bibitem{buchler} H.P. B\"{u}chler, V.B. Geshkenbein, and G. Blatter, Phys. Rev. Lett. \textbf{92}, 067007 (2004).

\bibitem{hekking_glazman} F.W.J. Hekking and L.I. Glazman, Phys. Rev. B \textbf{55}, 6551 (1997).

\bibitem{rob_qp} C. Wang, Y.Y. Gao, I.M. Pop, U. Vool, C. Axline, T. Brecht, R.W. Heeres, L. Frunzio, M.H. Devoret, G. Catelani, L.I. Glazman, and R.J. Schoelkopf, Nat. Commun. \textbf{5}, 5836 (2014).

\bibitem{michel_qp} K. Serniak, M. Hays, G. de Lange, S. Diamond, S. Shankar, L.D. Burkhart, L. Frunzio, M. Houzet, and M.H. Devoret, Phys. Rev. Lett. \textbf{121}, 157701 (2018).

\bibitem{kamenev_qp} Ya. Savich, L. Glazman, and A. Kamenev, Phys. Rev. B \textbf{96}, 104510 (2017).


\bibitem{roman} R.M. Lutchyn, J.D. Sau, and S. Das Sarma, Phys. Rev. Lett. \textbf{105}, 077001 (2010).

\bibitem{majorana_vonoppen} Y. Oreg, G. Refael, and F. von Oppen, Phys. Rev. Lett. \textbf{105}, 177002 (2010).

\bibitem{michel_rob_outlook} M.H. Devoret and R.J. Schoelkopf, Science \textbf{339},   1169 (2013).





\bibitem{newbower} R.S. Newbower, M.R. Beasley, and M. Tinkham, Phys. Rev. B \textbf{5}, 864 (1972).

\bibitem{lau_qps} C.N. Lau, N. Markovic, M. Bockrath, A. Bezryadin, and M. Tinkham, Phys. Rev. Lett. \textbf{87}, 217003 (2001).

\bibitem{bezryadin_qps} A. Bezryadin, C.N. Lau, and M. Tinkham, Nature(London) \textbf{404}, 971 (2000).

\bibitem{altomare} F. Altomare, A.M. Chang, M.R. Melloch, Y. Hong, and C.W. Tu, Phys. Rev. Lett. \textbf{97}, 017001 (2006).

\bibitem{zgirski} M. Zgirski, K.-P. Riikonen, V. Touboltsev, and K.Yu. Arutyunov, Phys. Rev. B \textbf{77}, 054508 (2008).

\bibitem{bollinger} A.T. Bollinger, A. Rogachev, and A. Bezryadin, Europhys. Lett. \textbf{76}, 505 (2006).

\bibitem{rogachev_1} A. Rogachev and A. Bezryadin, Appl. Phys. Lett. \textbf{83}, 512 (2003). 

\bibitem{rogachev_2} A. Rogachev, A.T. Bollinger, and A. Bezryadin, Phys. Rev. Lett. \textbf{94}, 017004 (2005). 






\bibitem{bezryadin_sahu} M. Sahu, M.-H. Bae, A. Rogachev, D. Pekker, T.-C. Wei, N. Shah, P. M. Goldbart, and A. Bezryadin, Nat. Phys. \textbf{5}, 503 (2009).

\bibitem{li} P. Li, P.M. Wu, Yu. Bomze, I.V. Borzenets, G. Finkelstein, and A.M. Chang, Phys. Rev. Lett. \textbf{107}, 137004 (2011).

\bibitem{aref} T. Aref, A. Levchenko, V. Vakaryuk, and A. Bezryadin, Phys. Rev. B \textbf{86}, 024507 (2012).



\bibitem{byers_yang} N. Byers  and C.N. Yang, Phys. Rev. Lett. \textbf{7}, 46 (1961).

\bibitem{little_parks2}  R.D. Parks,  and W.A. Little,  Phys. Rev. \textbf{133},  A97 (1964).



\bibitem{belkin} A. Belkin, M. Belkin, V. Vakaryuk, S. Khlebnikov, and A. Bezryadin, Phys. Rev. X \textbf{5}, 021023 (2015).


\bibitem{ps_rings_1} J.E. Lukens and J.M. Goodkind, Phys. Rev. Lett. \textbf{20}, 1363 (1968).

\bibitem{ps_rings_2} J.R. Kirtley, C.C. Tsuei, V.G. Kogan, J.R. Clem, H. Raffy, and Z.Z. Li, Phys. Rev. B \textbf{68}, 214505 (2003).

\bibitem{zhang_price} X. Zhang and J.C. Price,  Phys. Rev. B \textbf{55}, 3128 (1997).

\bibitem{budakian} H. Polshyn, T.R. Naibert, and R. Budakian, Phys. Rev. B \textbf{97}, 184501 (2018).





\bibitem{deterministic} I. Petkovic, A. Lollo, L.I. Glazman, and J.G.E. Harris, Nat. Commun. \textbf{7}, 13551 (2016).

\bibitem{will_thesis} W.E. Shanks, Persistent currents in normal metal rings. Ph.D. thesis, Yale University, 2011.




\bibitem{si} See Supplemental Material.

\bibitem{fulton_dunkelberger}  A. Fulton and L.N. Dunkleberger, Phys. Rev. B \textbf{9}, 4760 (1974).

\bibitem{garg} A. Garg, Phys. Rev. B \textbf{51}, 15592 (1995).

\bibitem{eckhaus} W. Eckhaus, \emph{Asymptotic Analysis II}, edited by F. Verhulst, Lecture Notes in Mathematics Vol 985 (Springer, New York, 1983), p. 449.  

\bibitem{khlebnikov} S. Khlebnikov, Phys. Rev. B \textbf{95}, 174507 (2017).


\bibitem{ambegaokar_halperin} V. Ambegaokar and B.I. Halperin, Phys. Rev. Lett. \textbf{22}, 1364 (1969). 

\bibitem{martinis_kautz} J.M. Martinis and R.L. Kautz, Phys. Rev. Lett. \textbf{63}, 1507 (1989).

\bibitem{michel1} P. Joyez, D. Vion, M. G\"{o}tz, M.H. Devoret, and D. Esteve, J. Supercond. \textbf{12}, 757 (1999). 

\bibitem{michel2} D. Vion, M. G\"{o}tz, P. Joyez, D. Esteve, and M.H. Devoret, Phys. Rev. Lett. \textbf{77}, 3435 (1996).

\bibitem{boogard} G.R. Boogaard, A.H. Verbruggen, W. Belzig, and T.M. Klapwijk, Phys. Rev. B \textbf{69}, 220503(R) (2004).



\end{thebibliography}

\begin{thebibliography}{99}

\bibitem{will_thesiss} W.E. Shanks, Persistent currents in normal metal rings. Ph.D. thesis, Yale University, 2011.

\bibitem{deterministicc}  I. Petkovic, A. Lollo, L.I. Glazman, and J.G.E. Harris, Nat. Commun. \textbf{7}, 13551 (2016).

\bibitem{zhang_pricee} X. Zhang and J.C. Price,  Phys. Rev. B \textbf{55}, 3128 (1997).

\bibitem{la} J.S Langer and V. Ambegaokar,  Phys. Rev. \textbf{164}, 498 (1967).



\end{thebibliography}
\end{document}